Linear Fractionally Damped Oscillator

Mark Naber [a]

Department of Mathematics

Monroe County Community College

Monroe, Michigan, 48161-9746

In this paper the linearly damped oscillator equation is considered with the damping term generalized to a Caputo fractional derivative. The order of the derivative being considered is $0 \leq \nu \leq 1$. At the lower end ($\nu = 0$) the equation represents an un-damped oscillator and at the upper end ($\nu = 1$) the ordinary linearly damped oscillator equation is recovered. A solution is found analytically and a comparison with the ordinary linearly damped oscillator is made. It is found that there are nine distinct cases as opposed to the usual three for the ordinary equation (damped, over-damped, and critically damped). For three of these cases it is shown that the frequency of oscillation actually increases with increasing damping order before eventually falling to the limiting value given by the ordinary damped oscillator equation. For the other six cases the behavior is as expected, the frequency of oscillation decreases with increasing order of the derivative (damping term).

[a] Electronic mail: mnaber@monroeccc.edu

# 1. INTRODUCTION

In this paper the linearly damped oscillator equation is considered with the damping term replaced by a fractional derivative [1] whose order, $v$, will be restricted to, $0 \leq v \leq 1$,

$$D_t^2 x + \lambda_0 D_t^v x + \omega^2 x = 0 . \tag{1}$$

Burov and Barki [2, 3] examined such an equation in connection with critical behavior. They were able to determine a solution in terms of generalized Mittag-Leffler functions. Non-linear fractional oscillators have been studied numerically by Zaslavsky [4]. He was primarily interested in chaotic behavior. It is hoped that a careful study of the analytic solution to the linear fractionally damped equation will help shed light on properties of the nonlinear equation and be of use for direct applications of fractionally damped oscillations (see for example [5, 6]).

In this paper the Caputo formulation of the fractional derivative will be used. The Caputo derivative is preferred over the Riemann-Liouville derivative for physical reasons. Consider the Laplace transform of the two formulations of the fractional derivative for $0 < v < 1$

$$\mathsf{L}\left({}_0^{RL}D_t^v f(t)\right) = s^v F(s) - {}_0^{RL}D_t^{v-1} f(t)\Big|_0, \tag{2}$$

$$\mathsf{L}\left({}_0^{C}D_t^v f(t)\right) = s^v F(s) - f(t)\Big|_0. \tag{3}$$

The constant term arising from the Laplace transform of the Caputo derivative is merely the initial value of the function. For the Riemann-Liouville derivative this is not the case. The constant term arising from the Laplace transform currently has no simple physical interpretation. Hence the Caputo fractional derivative seems to be more useful for modeling physical systems.

If the order of the fractional damping term is allowed to become 3/2 (outside the range of values considered in this paper) the equation is usually referred to as the Bagley-Torvik equation (see for example [1, 7, 8]). The solution of this equation exhibits damped oscillatory behavior similar to what we expect to find for the equation studied in this paper. The Bagley-Torvik equation was originally derived to study the motion of a rigid plate in a Newtonian fluid [7].

The analytic solution to the fractionally damped equation is found by means of Laplace transform. For the sake of clarity, and for pointing out some unique difficulties with the factional equation, a comparison with the Laplace transform method as applied to the non-fractional case is made. It is found that there are nine distinct cases for the fractionally damped equation as opposed to the usual three cases for the non-fractional equation. In six of the nine cases the results are as expected; increasing the order of the factional derivative increases the effects of the damping (i.e. the frequency of the damping slows as the order of the derivative increases). However, in three cases, the frequency of the damping actually increases as the order of the fractional derivative increases until a peak value is reached after which the frequency falls to its non-fractional limit. The physical reason for this increase in the oscillation frequency is not yet clear.

## 2. THE NON-FRACTIONAL CASE

Before a solution to the linear fractionally damped oscillator equation is constructed it will be useful to review the Laplace transform method of solution for the linearly damped oscillator equation,

$$D_t^2 x + \lambda D_t x + \omega^2 x = 0. \tag{4}$$

The constants $\lambda$ and $\omega$ are taken to be real and positive. $\lambda$ is the damping force per unit mass. $\omega^2$ is the restoring force per unit mass. In both cases (fractional and non-fractional) the following initial conditions will be used

$$\begin{aligned} x(0) &= x_0, \\ D_t x(0) &= x_1. \end{aligned} \tag{5}$$

Transforming Eq. (4) together with the initial conditions Eq. (5) gives the following

$$s^2 X(s) - s x_0 - x_1 + \lambda \left( s X(s) - x_0 \right) + \omega^2 X(s) = 0, \tag{6}$$

or,

$$X(s) = \frac{s x_0}{s^2 + \lambda s + \omega^2} + \frac{x_1 + \lambda x_0}{s^2 + \lambda s + \omega^2}. \tag{7}$$

Eq. (7) can be inverted using tables, however, to shed light on a problem that will happen later, Eq. (7) will be inverted via the complex inversion integral. The exponents on the $s$ variable in both terms are whole numbers, hence there will not be a branch cut in the contour integral and the Bromwich contour can be used,

$$x(t) = \text{Residue} - \frac{1}{2\pi i} \int e^{st} X(s) ds. \tag{8}$$

Recall that the Bromwich contour begins at $\gamma - i\infty$ goes vertically up to $\gamma + i\infty$ (where $\gamma$ is chosen so that all poles will lie to the left of the vertical contour line and thus all poles will be captured within the contour) and then travels in a half circle (to the left, counter clockwise) back to $\gamma - i\infty$. For this problem there is no contribution from the contour integral. The only contribution comes from the residue. The residue is generated from the roots of the following quadratic equation,

$$s^2 + \lambda s + \omega^2 = 0. \tag{9}$$

There are three different cases.

1) $\lambda > 2\omega$ ; 2 unequal real roots that are negative.

$$s_{1,2} = \frac{-\lambda \pm \sqrt{\lambda^2 - 4\omega^2}}{2} \tag{10}$$

2) $\lambda = 2\omega$ ; 2 repeated real roots that are negative.

$$s_3 = \frac{-\lambda}{2} = -\omega \tag{11}$$

3) $\lambda < 2\omega$ ; 2 complex roots whose real parts are negative.

$$s_{4,5} = \frac{-\lambda \pm i\sqrt{4\omega^2 - \lambda^2}}{2} \tag{12}$$

See fig. 1 below for a graphical representation of the location of the roots.

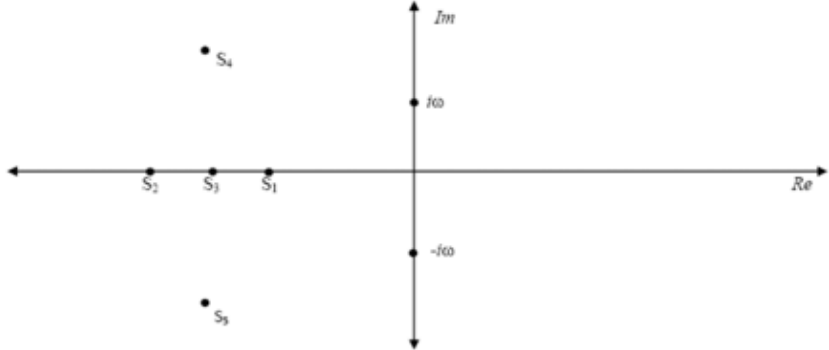

Fig. 1

Note that in cases one and three the poles will be of order one and in case two the pole will be of order two. Note also that if there were no damping the poles would be on the imaginary axis at $\pm i\omega$.

Computing the residue for case one gives

$$\text{Residue} = \lim_{s \to s_{1,2}} (s - s_{1,2}) e^{st} \left( \frac{sx_0}{s^2 + \lambda s + \omega^2} + \frac{x_1 + \lambda x_0}{s^2 + \lambda s + \omega^2} \right), \tag{13}$$

or,

$$x(t) = \frac{e^{s_1 t}}{2s_1 + \lambda}(s_1 x_0 + x_1 + \lambda x_0) + \frac{e^{s_2 t}}{2s_2 + \lambda}(s_2 x_0 + x_1 + \lambda x_0). \tag{14}$$

As $s_1$ and $s_2$ are both negative this solution will decay exponentially. This is usually referred to as the over-damped case.

Case three is computed the same way as case one. Now the poles are complex so the exponential function can be expressed using sine and cosine with an over-all exponential damping factor

$$x(t) = e^{-\alpha t}\left(x_0 \cos(\rho t) + \frac{(2x_1 + \lambda x_0)}{2\rho}\sin(\rho t)\right), \tag{15}$$

where $\rho = \sqrt{\omega^2 - \lambda^2/4}$ and $\alpha = \frac{\lambda}{2}$. Notice that the presence of damping causes the effective angular frequency, $\rho$, to be smaller than the un-damped angular frequency, i.e. the oscillations go slower, as one might expect if there were damping to impede the motion. This is usually referred to as the under-damped case. By comparison, cases one and two could be viewed as having a zero frequency or an infinite period.

For case two the pole is of order 2, and the residue is given by

$$\lim_{s \to -\omega} \frac{d}{ds}\left((s+\omega)^2 e^{st}\left(\frac{sx_0 + x_1 + \lambda x_0}{s^2 + \lambda s + \omega^2}\right)\right). \tag{16}$$

Recall that $\lambda = 2\omega$, this allows the denominator to be factored. The limit then becomes,

$$\lim_{s \to -\omega} \frac{d}{ds}\left(e^{st}(sx_0 + x_1 + 2\omega x_0)\right), \tag{17}$$

or,

$$x(t) = e^{-\omega t}\left(t(\omega x_0 + x_1) + x_0\right). \tag{18}$$

This is usually called the critically-damped case. Graphs of sample solutions to these three cases can be found in any introductory book on differential equations.

### 3. THE FRACTIONAL CASE

Now consider Eq. (1). In this case, $\lambda$ has units of time raised to the power $v - 2$. Hence the over all units of the second term remain the same as in Eq. (4). There are two cases to consider, $0 < v < 1$ and $1 < v < 2$. We will consider the former in this paper. The Laplace transform of Eq. (1) is

$$s^2 X - sx_0 - x_1 + \lambda\left(s^v X - s^{v-1} x_0\right) + \omega^2 X = 0, \tag{19}$$

or,

$$X = \frac{sx_0 + x_1 + \lambda s^{\nu-1} x_0}{s^2 + \lambda s^\nu + \omega^2}. \qquad (20)$$

If Eq. (20) is inverted using a contour integral a branch cut is needed on the negative real axis due to the fractional exponents on the complex variable $s$. Hence, a Hankel contour will be used. This contour starts at $\gamma - i\infty$ goes vertically up to $\gamma + i\infty$ (where $\gamma$ is again chosen so that all poles will lie to the left of the vertical contour line) and then travels in a quarter circle arc (to the left) to just above the negative real axis (i.e. $-\infty$). The contour then has a cut that goes into the origin (following the negative real axis), around the origin in a clockwise sense (to just below the negative real axis) and then back out to $-\infty$. The contour then has another quarter circle arc to $\gamma - i\infty$.

Now the question is, where are the poles? This is a somewhat more involved question than in the standard linearly damped model. To find the poles the following equation needs to be solved

$$s^2 + \lambda s^\nu + \omega^2 = 0. \qquad (21)$$

Which, for an arbitrary $\nu$, is not a trivial problem. To determine if there are solutions, and if so how many, let $s = re^{i\theta}$ then Eq. (21) breaks into 2 equations, a real and an imaginary part

$$r^2 \cos(2\theta) + \lambda r^\nu \cos(\nu\theta) + \omega^2 = 0,$$
$$r^2 \sin(2\theta) + \lambda r^\nu \sin(\nu\theta) = 0. \qquad (22)$$

Could there be a solution on the positive real axis? No, in this case $\theta = 0$ and the first equation of Eq. (22) would be the sum of three positive non-zero terms, which would never be zero. Could there be a solution on the negative real axis? No, in this case $\theta = \pi$ and the second term of the second equation of Eq. (22) would never be zero. Using similar arguments we can show that there are no solutions on the positive or negative imaginary axes, recall $0 < \nu < 1$. It can also be shown that no solutions are in the right half plane (both terms of the second equation would always be positive). If there are solutions they should be in pairs, complex conjugates, with $\pi/2 < \theta < \pi$ and $-\pi/2 > \theta > -\pi$. To attempt to find a solution first solve the second equation of Eq. (22) for $r$ and substitute this into the first equation (only look for positive $\theta$ values first),

$$\left(-\lambda \frac{\sin(\nu\theta)}{\sin(2\theta)}\right)^{2/(2-\nu)} \cos(2\theta) + \lambda \left(-\lambda \frac{\sin(\nu\theta)}{\sin(2\theta)}\right)^{\nu/(2-\nu)} \cos(\nu\theta) + \omega^2 = 0. \qquad (23)$$

The reader may be worried about the negative sign and the fractional exponent in Eq. (23), however, for the restricted angular range being considered, $\pi/2 < \theta < \pi$, $\sin(2\theta)$ is always negative. So, the argument of the root will always be positive.

Given values for $v$, $\lambda$, and $\omega$ it would appear to be impossible to solve Eq. (23) for $\theta$. Equation (23) can be simplified to a more aesthetically pleasing form,

$$\left(\frac{(\sin(v\theta))^v}{(\sin(2\theta))^2}\right)^{1/(2-v)} \sin((2-v)\theta) = \left(\frac{\omega}{\lambda^{1/(2-v)}}\right)^2. \tag{24}$$

Now it needs to be seen if there is a $\theta$ value that will satisfy Eq. (24). For this equation to be true $\sin((2-v)\theta)$ needs to be positive. This will only happen on the restricted domain $\frac{\pi}{2} < \theta < \frac{\pi}{2-v}$. Now the question becomes, on this restricted domain can we pick a $\theta$ value that will make the left hand side of Eq. (24) as large or as small as we wish? Thus ensuring that no matter what the values of $\omega$, $\lambda$, and $v$ we are given we can always find a $\theta$ value that will satisfy Eq. (24). Consider the two limits

$$\lim_{\theta \to \pi/2^+} \left(\frac{(\sin(v\theta))^v}{(\sin(2\theta))^2}\right)^{1/(2-v)} \sin((2-v)\theta) = \infty, \tag{25}$$

$$\lim_{\theta \to \pi/(2-v)^-} \left(\frac{(\sin(v\theta))^v}{(\sin(2\theta))^2}\right)^{1/(2-v)} \sin((2-v)\theta) = 0. \tag{26}$$

Since the left hand side of Eq. (24) is continuous in $\theta$ and we have the two limits above, Eqs. (25) and (26), it is guaranteed that there will be at least one solution to Eq. (24) and hence there will be at least two poles for the residue calculation. If we can show that the left hand side of Eq. (24) decreases monotonically in $\theta$ over the restricted domain then we know that there will be only one solution to Eq. (24), and thus only two poles in the residue calculation. To show that the left hand side of Eq. (24) decreases monotonically in $\theta$ we need to show that the derivative of the left hand side of Eq. (24) with respect to $\theta$ is always negative, i.e.

$$\frac{\partial}{\partial \theta}\left\{\left(\frac{(\sin(v\theta))^v}{(\sin(2\theta))^2}\right)^{1/(2-v)} \sin((2-v)\theta)\right\} < 0. \tag{27}$$

Computing the derivative, doing some algebra, and throwing away over all factors that are always positive we have,

$$v^2 \sin^2(2\theta) - 4v \sin(2\theta)\sin(v\theta)\cos((2-v)\theta) + 4\sin^2(v\theta) > 0. \tag{28}$$

On the restricted domain $\sin(v\theta) > 0$, $\sin(2\theta) < 0$, and $\cos((2-v)\theta) \leq 1$. This reduces Eq. (28) to

$$v^2 \sin^2(2\theta) - 4v\sin(2\theta)\sin(v\theta) + 4\sin^2(v\theta) > 0. \tag{29}$$

Eq. (29) can now be factored into a perfect square and prove the assertion made in Eq. (27),

$$(v\sin(2\theta) - 2\sin(v\theta))^2 > 0. \tag{30}$$

Hence the left hand side of Eq. (24) will decrease monotonically on the restricted domain with the upper bound being $\infty$ and the lower bound being 0. To summarize, it has just been shown that there is always one solution, with a positive angle, to Eq. (24) and this solution must be such that $\frac{\pi}{2} < \theta < \frac{\pi}{2-v}$. Consequently there will be two poles for the residue calculation and they will be complex conjugates of each other. Notice that for the fractionally damped equation repeated roots are not possible. Repeated roots can only happen when the order of the derivative becomes one. See fig. 2 below for a graphical representation of the location of the roots.

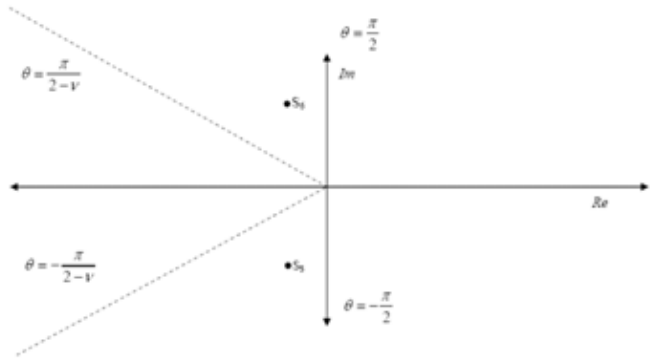

Fig. 2

Now that the question of the poles has been settled the solution to Eq. (1) can be generated. Denote the two poles as

$$s_{6,7} = \beta \pm i\sigma = re^{\pm i\theta}, \tag{31}$$

where $\beta$ and $\sigma$ are determined from the $r$ and $\theta$ values that satisfy Eq. (24) in the usual way, $r = \sqrt{\beta^2 + \sigma^2}$ and $\tan(\theta) = \sigma/\beta$. Note that $\beta$ is negative, the two solutions are in the second and third quadrants and $s_7$ is the complex conjugate of $s_6$. Note also that $\beta$ plays the role of $-\lambda/2$ from the non-fractional case. The poles are of order one and the residue is given by,

$$\text{Residue} = \lim_{s \to s_6}(s - s_6)e^{st}\left(\frac{sx_0 + x_1 + \lambda s^{v-1}x_0}{s^2 + \lambda s^v + \omega^2}\right) + \lim_{s \to s_7}(s - s_7)e^{st}\left(\frac{sx_0 + x_1 + \lambda s^{v-1}x_0}{s^2 + \lambda s^v + \omega^2}\right), \tag{32}$$

$$= e^{s_6 t} \left( \frac{s_6 x_0 + x_1 + x_0 \lambda s_6^{\nu-1}}{2 s_6 + \nu \lambda s_6^{\nu-1}} \right) + e^{\bar{s}_6 t} \left( \frac{\bar{s}_6 x_0 + x_1 + x_0 \lambda \bar{s}_6^{\nu-1}}{2 \bar{s}_6 + \nu \lambda \bar{s}_6^{\nu-1}} \right), \tag{33}$$

where $s_7$ has been replaced by $\bar{s}_6$. After some algebra this can be reduced to,

$$2 e^{\beta t} \cos(\sigma t) \frac{x_0 \left( 2r^2 + \nu \lambda^2 r^{2\nu-2} + \lambda r^\nu (\nu + 2) \cos(\theta(\nu-2)) \right) + x_1 \left( 2r \cos(\theta) + \nu \lambda r^{\nu-1} \cos(\theta(\nu-1)) \right)}{4r^2 + 4\nu \lambda^\nu r^\nu \cos((2-\nu)\theta) + \nu^2 \lambda^{2\nu} r^{2\nu-2}}$$

$$+2 e^{\beta t} \sin(\sigma t) \frac{x_0 \left( \lambda r^\nu (\nu - 2) \sin((\nu-2)\theta) \right) + x_1 \left( 2r \sin(\theta) + \nu \lambda r^{\nu-1} \sin(\theta(\nu-1)) \right)}{4r^2 + 4\nu \lambda r^\nu \cos((2-\nu)\theta) + \nu^2 \lambda^2 r^{2\nu-2}}$$

(34)

For the contour integral the only contributions come from the paths along the negative real axis

$$\frac{\lambda}{\pi} \int_0^\infty \frac{(Rx_0 - x_1) \sin(\nu\pi) + \frac{x_0}{R}(R^2 + \omega^2) \sin(\pi(\nu-1))}{(R^2 + \omega^2)^2 + 2\lambda R^\nu (R^2 + \omega^2) \cos(\nu\pi) + (\lambda R^\nu)^2} e^{-Rt} R^\nu dR. \tag{35}$$

The solution to Eq. (1) is then, Eq. (35) subtracted from Eq. (34). This may look overly complicated but the solution does have the general form of

$$x(t) = A e^{\beta t} \cos(\sigma t) + B e^{\beta t} \sin(\sigma t) - \text{Decay function}. \tag{36}$$

Notice that the decay function, Eq. (35), goes to zero if $\nu$ goes to zero or one, i.e. if Eq. (1) goes to its non-fractional limits the decay function goes away, as expected. The damping factor $e^{\beta t}$ is similar to the damping factor for the non-fractional case, $e^{-\lambda t/2}$. Notice that since the poles, for the residue calculation, have non-zero imaginary and non-zero real parts we will not have the same three distinct cases as we did for the non-fractional case (critically-damped, over-damped, and under-damped).

## 4. THE OSCILLATION FREQUENCY

Consider the frequency of the oscillation component of the solution, $\sigma = \text{Im}(s_6)$. One question we might ask is: how does the frequency change as we change the order of the fractional damping? When $\nu$ is set to zero we have an un-damped oscillator with frequency

$$\sigma = \sqrt{\lambda + \omega^2}. \tag{37}$$

When $v$ is set to one we have the three cases given at the beginning of the chapter; over-damped, critically-damped, and under-damped. So, the frequency may be zero or non-zero, i.e.,

$$\sigma = 0 \qquad \lambda \geq 2\omega,$$
$$\sigma = \frac{\sqrt{4\omega^2 - \lambda^2}}{2} \qquad \lambda < 2\omega. \tag{38}$$

For $0 < v < 1$ there will always be a non-zero frequency. Note that $0 \leq \sqrt{4\omega^2 - \lambda^2}/2 < \sqrt{\lambda + \omega^2}$.

In the non-fractional case increasing $\lambda$ causes the frequency of oscillation to become smaller, monotonically, until the critical cases are reached and the oscillation period becomes infinite (these are the critical and over-damped cases). In the fractional case the frequency of oscillation, $\sigma = \text{Im}(s_6)$, now depends on the order of the derivative, $v$, as well as $\lambda$ and $\omega$. To try to determine how $\sigma$ depends on these three parameters consider $s$ to be a function of $v$, on $0 \leq v \leq 1$, implicitly defined by

$$s^2 + \lambda s^v + \omega^2 = 0, \tag{39}$$

for fixed values of $\lambda$ and $\omega$ (both being positive). Let us restrict our attention to the upper half plane for $s$. As such $s$ will be one-to-one on $0 \leq v < 1$. Due to the fractional exponent causing a branch cut on the negative real axis $s$ will not be one-to-one at $v = 1$.

Now the question arises, does $\sigma$ fall monotonically with respect to $v$? To get at the answer to this consider the derivative of Eq. (39) with respect to $v$ and isolate $ds/dv$ (remember, $\lambda$ and $\omega$ are being held fixed)

$$\frac{ds}{dv} = -\frac{\lambda s^v \ln(s) s}{2s^2 + \lambda s^v v}. \tag{40}$$

The imaginary part of this equation is,

$$\frac{d\sigma}{dv} = \text{Im}\left(\frac{ds}{dv}\right). \tag{41}$$

Specifically, consider this equation at $v = 0$

$$\left.\frac{d\sigma}{dv}\right|_{v=0} = \text{Im}\left(\frac{(s^2 + \omega^2)\ln(s)s}{2s^2 - v(s^2 + \omega^2)}\right)\bigg|_{v=0} = \frac{\lambda\left(\ln(\lambda + \omega^2)\right)}{4\sqrt{\lambda + \omega^2}}. \tag{42}$$

This gives three initial slopes for the rate of change of $\sigma$ with respect to $v$.

$\lambda + \omega^2 > 1 \implies$ The frequency initially increases with increasing damping order.

$\lambda + \omega^2 = 1 \implies$ The frequency initially is not changing with increasing damping order.

$\lambda + \omega^2 < 1 \implies$ The frequency initially decreases with increasing damping order.

This is not entirely what might have been expected. In the first case the oscillation frequency actually increases before falling. Hence there will be some values of $\nu$ for which the fractional damping will actually cause the oscillations to go faster than the un-damped oscillator (the damping will still cause the amplitude to decrease). Each of the above three cases can become any of the three non-fractionally damped cases by letting $\nu \to 1$ (Eqs. (10), (11), and (12)). Hence, there are nine cases for the linear fractionally damped oscillator.

Below are some graphs of solutions to the imaginary part of Eq. (39) (the oscillation frequency) for various values of $\nu$, $\lambda$, and $\omega$. In all three graphs the oscillation frequency is on the vertical axis and the order of the derivative is on the horizontal axis. The three graphs for each case correspond to what would be under-damped, critically-damped, and over-damped for a damped oscillator with whole order derivatives.

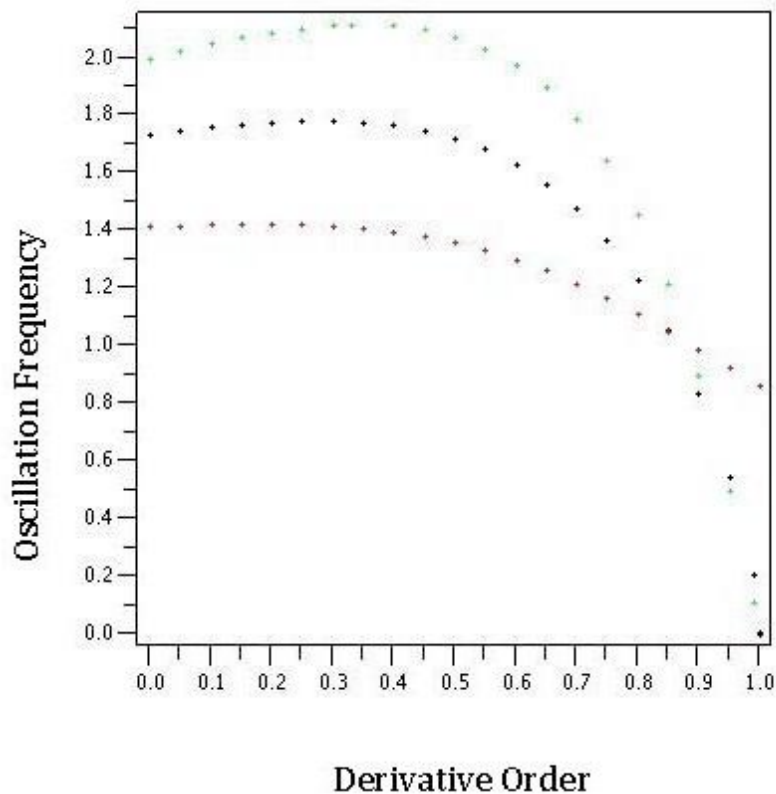

Fig. 3

Fig. 3 is a representative graph of case one, $\lambda + \omega^2 > 1$. The green graph is for $\lambda = \omega = 1$, the black graph is for $\lambda = 2$ and $\omega = 1$, and the red graph is for $\lambda = 3$ and $\omega = 1$.

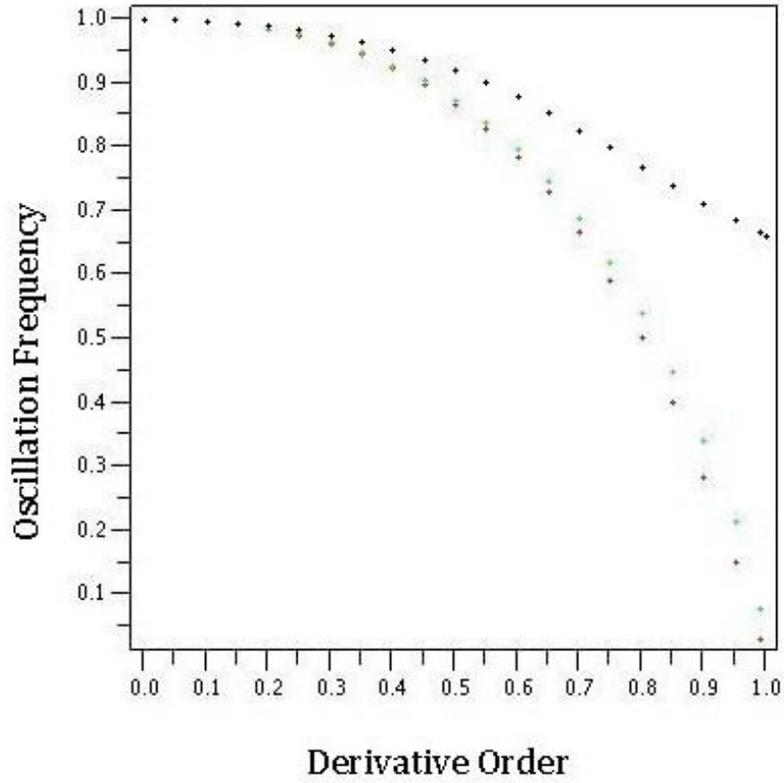

Fig. 4

Fig. 4 is a representative graph from case two, $\lambda + \omega^2 = 1$, a flat start. The red graph is for $\lambda = 2(\sqrt{2} - 1)$ and $\omega = \lambda/2$, the green graph is for $\lambda = 1/2$ and $\omega = 1/\sqrt{2}$, and the black graph is for $\lambda = 15/16$ and $\omega = 1/4$.

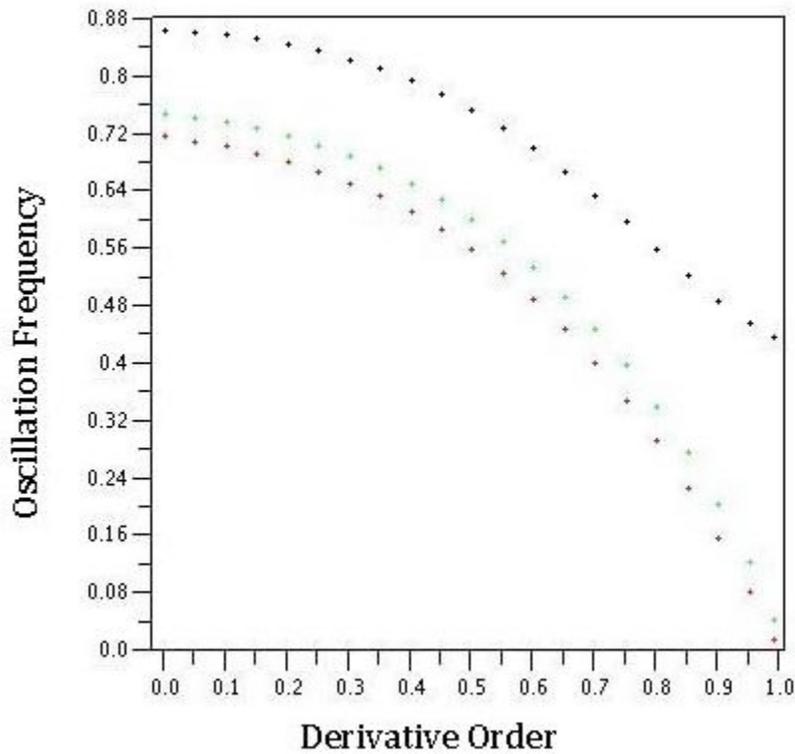

Fig. 5

Fig. 5 is a representative graph for case three a decreasing start. The green graph is for $\lambda = 1/2$ and $\omega = 1/8$, the black graph is for $\lambda = 1/2$ and $\omega = 1/4$, and the red graph is for $\lambda = \omega = 1/2$.

## 5. CONCLUSION

In this paper the linear fractionally damped oscillator equation was solved analytically. It was found that the solution is very similar to the non-fractional case (decayed oscillations but with the inclusion of an additional decay function). It was found that there are nine distinct cases, as opposed to the usual three for the ordinary damped oscillator. An unexpected result was that for three of the cases the oscillation frequency actually increases with increasing order of derivative of the damping term (till a peak value is reached, then the frequency decreases as expected). The physical reason for this increase in oscillation frequency is not yet clear.

## REFERENCES

[1] Igor Podlubny, *Fractional Differential Equations*, Academic Press, 1999.

[2] S. Burov and E. Barkai, Fractional Langevin Equation: Over-Damped, Under-Damped, and Critical behaviors, arXiv:0802.3777v1 [cond-mat.stat-mech] 26 Feb 2008.

[3] S. Burov and E. Barkai, The Critical Exponent of the Fractional Langevin Equation is $\alpha_c \approx 0.402$, arXiv:0712.3407v1 [cond-mat.stat-mech] 20 Dec 2007.

[4] G. M. Zaslavsky, A.A. Stanislavsky, and M. Edelman, *Chaotic and Pseudochaotic Attractors of Perturbed Fractional Oscillator*, arXiv:nlin.CD/0508018 v1 10 Aug 2005.

[5] Ana Cristina Galucio, Jean-François, and François Dubois, *On the use of fractional derivative operators to describe viscoelastic damping in structural dynamics- FE formulation of sandwich beams and approximation of fractional derivatives by using the $G^\alpha$ scheme*, Derivation fractionaire en mecanique – Etat-de-l'art et applications, CNAM Paris – 17[th] November, 2006.
http://www.cnam.fr/lmssc/seminaires/derivfrac/galucio/17NOV.PDF

[6] B. N. Narahari Achar, J.W. Hanneken, and T. Clarke, *Damping characteristics of a fractional oscillator*, Physica A: Statistical Mechanics and its Applications, Vol. 339, issues 3-4, 15 Aug 2004, pages 311-319.

[7] R. L. Bagley and P. J. Torvik, On the appearance of the fractional derivative in the behavior of real materials, J. Appl. Mech., 51 (1984), pp. 294-298.

[8] S. Saha Ray and R. K. Bera, Analytical solutions of the Bagley Torvik equation by Adomian decomposition method, Appl. Math. and Comp., vol 168, issue 1, Sept. 2005, pp. 398-410.